\documentclass[twocolumn,showpacs,preprintnumbers,prl,superscriptaddress]{revtex4}
\usepackage[dvipdfm]{graphicx}
\usepackage{amssymb}
\usepackage{amsmath}
\usepackage{epstopdf}
\usepackage{subfigure}
\usepackage{epsfig}
\usepackage{graphicx}
\usepackage[dvips]{color}

\makeatletter
\newcommand{\figcaption}[1]{\def\@captype{figure}\caption{#1}}
\newcommand{\tblcaption}[1]{\def\@captype{table}\caption{#1}}
\newcommand{\Nf}{N_{\rm f}}
\newcommand{\kh}{\kappa_{\rm h}}
\newcommand{\mh}{m_{\rm h}}
\newcommand{\ml}{m_{\rm l}}
\makeatother

\def\simge{\mathrel{%
       \rlap{\raise 0.511ex \hbox{$>$}}{\lower 0.511ex \hbox{$\sim$}}}}
\def\simle{\mathrel{
       \rlap{\raise 0.511ex \hbox{$<$}}{\lower 0.511ex \hbox{$\sim$}}}}

\begin{document}

\title{End Point of a First-Order Phase Transition in Many-Flavor Lattice QCD at Finite Temperature and Density}
\author{Shinji Ejiri}
\affiliation{
Graduate School of Science and Technology, Niigata University, Niigata 950-2181, Japan
}
\author{Norikazu Yamada}
\affiliation{
KEK Theory Center, Institute of Particle and Nuclear Studies, High
Energy Accelerator Research Organization (KEK), Tsukuba 305-0801,
Japan
}
\affiliation{
School of High Energy Accelerator Science, The Graduate University
for Advanced Studies (Sokendai), Tsukuba 305-0801, Japan
}

\date{March 30, 2013}

\begin{abstract}
Towards the feasibility study of the electroweak baryogenesis in
realistic technicolor scenario, 
we investigate the phase structure of $(2+\Nf)$-flavor QCD, where the mass
of two flavors is fixed to a small value and the others are heavy.
For the baryogenesis, an appearance of a first order phase transition at 
finite temperature is a necessary condition.
Using a set of configurations of two-flavor lattice QCD and applying
the reweighting method, the effective
potential defined by the probability distribution function of the
plaquette is calculated in the presence of additional many heavy flavors.
Through the shape of the effective potential, we determine the
critical mass of heavy flavors separating the first order and crossover 
regions and find it to become larger with $\Nf$.
We moreover study the critical line at finite density and the first
order region is found to become wider as increasing the chemical potential.
Possible applications to real $(2+1)$-flavor QCD are discussed.
\end{abstract}


\maketitle

\paragraph{\bf Introduction.}
Precise knowledge of the phase structure of finite temperature QCD could
offer
an opportunity to probe the physics beyond the standard model (SM),
provided that new gauge theory induces dynamical electroweak (EW)
symmetry breaking.
Technicolor (TC) is such a model~\cite{Weinberg:1975gm}, where the Higgs
sector in the SM is replaced by a new strongly interacting gauge theory
and its spontaneous chiral symmetry breaking (S$\chi$SB) causes
EW symmetry breaking.
TC is a vectorlike gauge theory, and if we choose SU(3) as a gauge
group it is essentially QCD.
The difference is only in their dynamical scales: $\sim$1 TeV for TC and
$\sim$1 GeV for QCD.
Thus, numerical techniques developed in lattice QCD trivially apply to
the study of TC, where the lattice cutoff is determined by equating
the pion decay constant to 246 GeV (Higgs vacuum expectation value).
The realizability of this model is now actively investigated using
lattice gauge theory~\cite{DelDebbio:2010zz}.
We consider TC including many fermion flavors transforming as the
fundamental representation of SU(3) since the presence of many flavors
potentially resolves various problems in classical TC.

In this work, we focus on the possibility of the EW baryogenesis within
the TC scenario~\cite{Appelquist:1995en}, which requires TC gauge theories
to go through a strong first order chiral phase transition.
The nature of the phase transition depends on the number of flavors and
masses~\cite{Pisarski:1983ms}.
In realistic TC models, two flavors of them are exactly massless and
the resulting three massless Nambu-Goldstone (NG) bosons are absorbed
into the longitudinal mode of the weak gauge bosons.
On the other hand, the mass of other $\Nf$ flavors must be larger than
an appropriate lower bound otherwise S$\chi$SB produces too many (light
pseudo) NG bosons, none of which is observed yet.
Consulting the study of (2+1)-flavor QCD including up, down and massive
strange quarks, the first order transition is realized when the strange
mass is below the critical mass.
Thus, requiring the first order EW phase transition in TC model
brings in the upper bound on the mass of $\Nf$ flavors.
This motivates us to study the thermal nature of $(2+\Nf)$-flavor QCD.
As discussed below, the critical mass increases with $\Nf$.
Hence, the boundary of the first order region can be investigated more
easily for large $\Nf$.

Another purpose of this study is to understand the real QCD with 2+1
flavors.
At the physical masses and zero density,
the chiral transition is crossover, and is expected to become first
order at a critical density.
The determination of the critical density is one of the most interesting
topics in the study of QCD. 
To this end, finding the critical surface in the masses and chemical
potential parameter space is important~\cite{dFP,crtpt}.
However, recent lattice QCD studies suggest that the critical region at 
zero density is accessible only when the quark masses are very small   
and thus its determination is difficult \cite{RBCBi09}.
The study of many-flavor QCD is a good 
testing ground for investigating $\Nf$-independent universal properties, 
such as the critical scaling near the tricritical point, which is expected 
in the up down quark massless limit.
This will provide important information for $(2+1)$-flavor QCD.

\paragraph{\bf Method.}
To study the phase transition, we calculate the effective potential
defined by the probability distribution function of the plaquette.
The distribution function has two peaks at a first order transition, 
since two phases coexist with the same probability.
The nature of the transition can be thus identified through the shape 
of the potential \cite{whot11,Ejiri:2007ga}.
We define the plaquette distribution 
function for $(2+\Nf)$-flavor QCD with the quark masses $m_f$ and 
chemical potential $\mu_f$
($f=1, \cdots, \Nf+2$) by
\begin{eqnarray}
w(P; \beta, m_f, \mu_f) 
&=&  \int {\cal D} U {\cal D} \psi {\cal D} \bar{\psi} \
    \delta(P- \hat{P}) \ e^{- S_q - S_g} \nonumber \\
&& \hspace{-30mm} = \int {\cal D} U \ \delta(P- \hat{P}) \ 
    e^{6\beta N_{\rm site} \hat{P}}\
    \prod_{f=1}^{\Nf+2} (\det M(m_f, \mu_f)),
\label{eq:pdist}
\end{eqnarray}
where 
$S_g$ and $S_q$ are the gauge and quark actions, respectively, and $M$
is the quark matrix.
$N_{\rm site} \equiv N_{\rm s}^3 \times N_t$ is the number of sites.
$\beta=6/g_0^2$ is the lattice bare parameter. 
$\hat P$ is the generalized plaquette operator, and
this method is applicable to the case of improved actions
replacing $\hat P$ to $\hat P=-S_g/(6N_{\rm site} \beta)$.
Normalizing  by the partition function, ${\cal Z}\!= \!\int\! w(P) dP$, 
eq.~(\ref{eq:pdist}) gives the histogram for $\hat P$.
The effective potential is then given by
\begin{eqnarray}
 V_{\rm eff}(P;\beta,m_f,\mu_f) = -\ln w(P;\beta,m_f,\mu_f).
 \label{eq:effective-potential}
\end{eqnarray}

We consider QCD with two degenerate light quarks of the mass
$m_{\rm l}$ and the chemical potential $\mu$ and $\Nf$ heavy quarks.
Denoting the potential of two-flavor QCD at $\mu$=0 by
$V_0(P; \beta)$, 
that of $(2+\Nf)$-flavor QCD
is written as
\begin{eqnarray}
    V_{\rm eff}(P; \beta, m_f, \mu)
=   V_0(P; \beta_0)
  - \ln R(P; \beta, m_f, \mu; \beta_0),
\label{eq:vefftrans}
\end{eqnarray}
with
\begin{eqnarray}
\ln R(P; \beta, m_f, \mu; \beta_0)
&=& 6(\beta - \beta_0)N_{\rm site}P 
\nonumber \\
& & \hspace{-38mm}
 + \ln
     \left\langle
     \displaystyle
     \left( \frac{\det M(\ml,\mu)}{\det M(\ml   ,0)}
     \right)^{\!\! 2} \prod_{f=1}^{\Nf}
     \frac{\det M(m_f,  \mu_f)}{\det M(\infty,0)}
     \right\rangle_{P: {\rm fixed}}, \ 
\label{eq:lnr}
\end{eqnarray}
where 
$ \langle \cdots \rangle_{P: {\rm fixed}} \equiv 
\langle \delta(P- \hat{P}) \cdots \rangle_{\beta_0} /
\langle \delta(P- \hat{P}) \rangle_{\beta_0} $
and $\langle \cdots \rangle_{\beta_0}$ denotes the ensemble average over
two-flavor configurations generated at $\beta_0$, $\ml$, and
vanishing $\mu$.
Since the $\ml$ dependence is not discussed in the
following, $\ml$ is omitted from the arguments.
$\beta_0$ is the simulation point, which may differ 
from $\beta$ in this method.
By performing simulations at various $\beta_0$, 
one can obtain the potential in a wide range of $P$.

Restricting the calculation to the heavy quark region,
the second determinant for $\Nf$ flavors in eq.~(\ref{eq:lnr}) is approximated
by the leading order as
\begin{eqnarray}
\ln \left[ \frac{\det M (\kh)}{\det M (0)} \right]  
=  288 N_{\rm site} \kh^4 \hat{P} + 12 N_s^3 (2 \kh)^{N_t} \hat{\Omega} 
+ \! \! \cdots 
\label{eq:detmw}
\end{eqnarray}
for the standard Wilson quark action and
\begin{eqnarray}
\ln \left[ \frac{\det M (m_{\rm h})}{\det M (\infty)} \right]  
=  \frac{36 N_{\rm site}}{(2m_{\rm h})^4} \hat{P} 
+  \frac{6 N_s^3}{(2m_{\rm h})^{N_t}} \hat{\Omega} 
+ \cdots 
\label{eq:detms}
\end{eqnarray}
for the four-flavor standard staggered quark with $m_{\rm h}$.
$\kh$ in eq.~(\ref{eq:detmw}) is the hopping parameter being 
proportional to $1/\mh$, and $\hat{\Omega}$ is the real part of the
Polyakov loop $\hat{\Omega}_R$ for $\mu_f=0$ and 
$\hat{\Omega} = \cosh(\mu_h /T) \hat{\Omega}_R 
+i \sinh(\mu_h /T) \hat{\Omega}_I$ 
for $\mu_f=\mu_{\rm h}$, including the complex phase 
from the imaginary part of the Polyakov loop $\hat{\Omega}_I$.
For improved gauge actions such as 
$S_g = -6N_{\rm site} \beta [c_0 {\rm (plaquette)} 
+ c_1 {\rm (rectangle)}]$, 
additional $c_1\times O(\kappa^4)$ terms must be contained 
in eqs.~(\ref{eq:detmw}) and (\ref{eq:detms}), 
where $c_1$ is the improvement coefficient and $c_0=1-8c_1$. 
However, since the improvement term does not affect the physics, 
we will cancel these terms by a shift of the coefficient $c_1$.

At a first order transition point, $V_{\rm eff}$ shows 
a double-well shape as a function of $P$, 
and, equivalently, the curvature of the potential $d^2 V_{\rm eff}/d^2P$ 
takes a negative value in a region of $P$.
To observe this behavior, $\beta$ must be adjusted to be the first 
order transition point.
However, from eqs.~(\ref{eq:vefftrans}) and (\ref{eq:lnr}), 
$d^2V_{\rm eff}/dP^2$ is independent of $\beta$.
The fine tuning is not necessary in this case \cite{Ejiri:2007ga}.
Moreover, $d^2V_{\rm eff}/dP^2$ over the wide range of $P$ can be easily
obtained by combining data obtained at different $\beta$.
We therefore focus on the curvature of
the effective potential to identify the nature of the phase transition.

Denoting $h=2 \Nf (2\kh)^{N_t}$ for $\Nf$ degenerate Wilson quarks, 
or $h=\Nf/(4 \times (2m_{\rm h})^{N_t})$ for the staggered quarks, 
we obtain
$\ln R(P;\beta,\kh,0;\beta_0)
=\ln\bar{R}(P; h,0)+{\rm (plaquette \ term)} + O(\kh^{N_t+2})$
for $\mu=\mu_{\rm h}=0$ with
\begin{eqnarray}
\bar{R}(P; h,0)
= \left\langle \exp [6h N_s^3 \hat{\Omega}] 
\right\rangle_{P: {\rm fixed}, \beta_0} .
\label{eq:rew2f}
\end{eqnarray}
Notice that $\bar{R}(P; h,0)$ does not depend on $\beta_0$.
The plaquette term does not contribute to $d^2V_{\rm eff}/dP^2$
and can be absorbed by shifting  
$\beta \to \beta^{*} \equiv \beta + 48 \Nf \kh^4$ for Wilson quarks. 
Moreover, one can deal with the case with non-degenerate masses by adopting
$h=2 \sum_{f=1}^{\Nf} (2 \kappa_f)^{N_t}$ or 
$h=(1/4)\sum_{f=1}^{\Nf} (2m_f)^{-N_t}$.
Thus, the choice of the quark action is not important.
In the following, we discuss the mass dependence of $\bar{R}$ 
through the parameter $h$.

\begin{figure}[tb]
\begin{center}
\begin{tabular}{c}
\includegraphics[width=71mm]{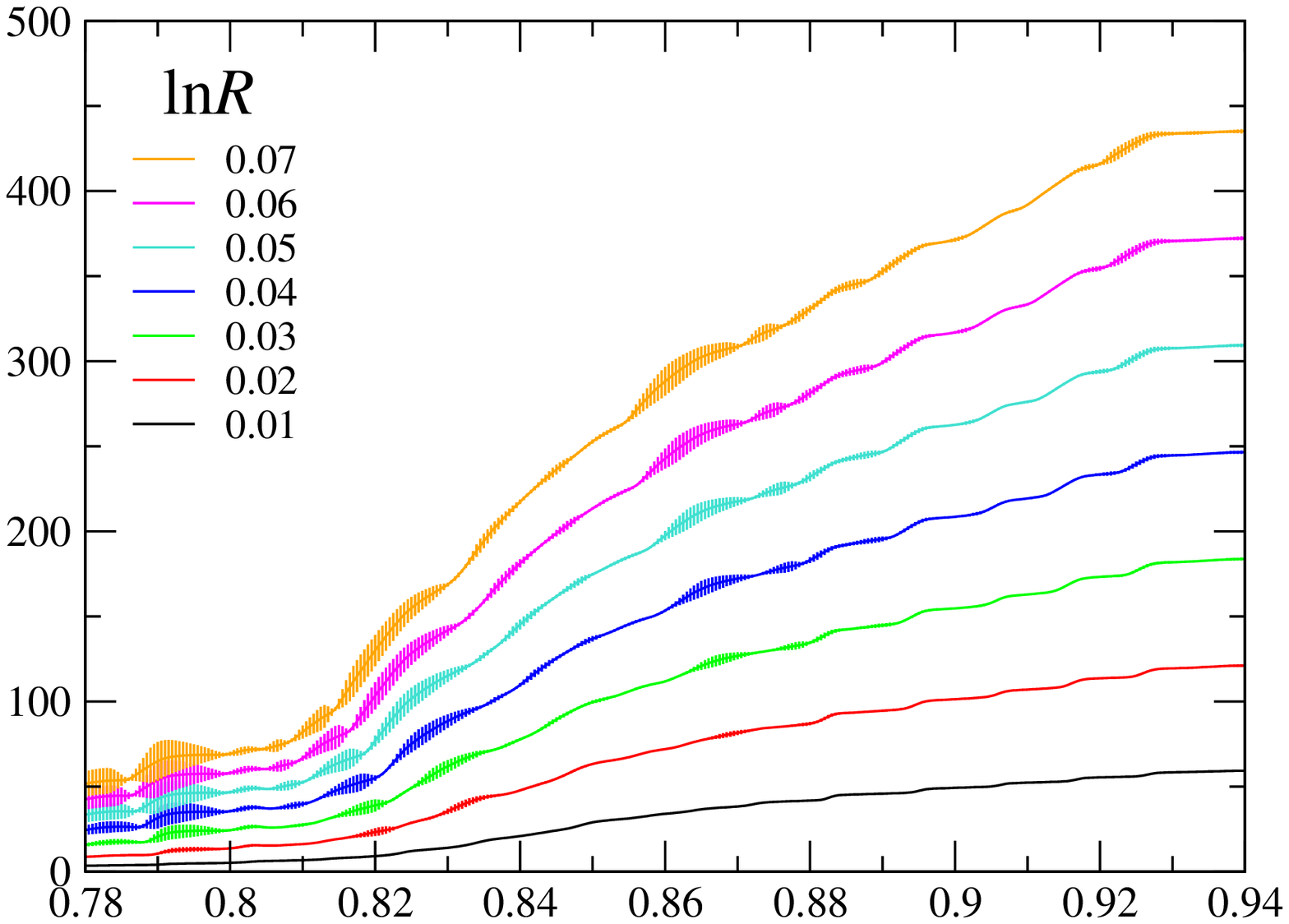} \\
\includegraphics[width=71mm]{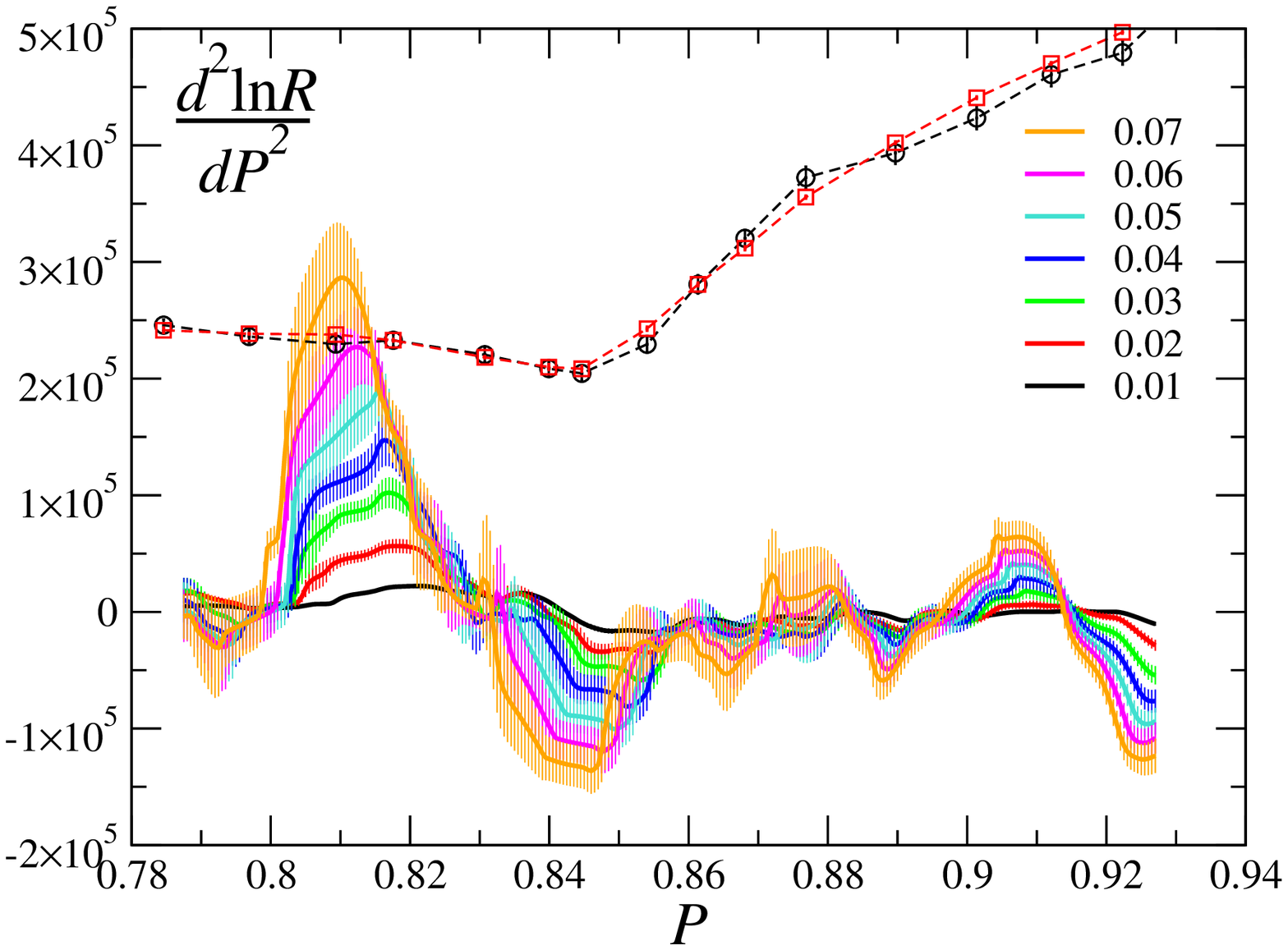}
\end{tabular}
\vspace{-3mm}
\caption{Top: $\ln \bar{R} (P; h,0)$ as functions of the plaquette.
Bottom: The curvature of $\ln \bar{R} (P; h,0)$ 
for $h=0.01$ -- $0.07$. 
The circle and square symbols are $d^2V_0/dP^2(P)$.}
\label{fig:lnr}
\end{center}
\end{figure}

\paragraph{\bf Numerical results.}
We use the two-flavor QCD configurations generated with p4-improved
staggered quark and Symanzik-improved gauge actions in
Ref.~\cite{BS05},
thus $\hat P=-S_g/(6N_{\rm site} \beta)$.
The lattice size $N_{\rm site}$ is $16^3 \times 4$. 
The data are obtained at sixteen values of $\beta$ from $\beta=3.52$ to
$4.00$ keeping the bare quark mass to $ma=0.1$.
The number of trajectories is 10000 -- 40000, depending on $\beta$.
The corresponding temperature normalized by the pseudo-critical
temperature is in the range of $T/T_c= 0.76$ to $1.98$, and the
pseudo-critical point is about $\beta=3.65$, where the ratio of
pseudo-scalar and vector meson masses is
$m_{\rm PS}/m_{\rm V} \approx 0.7$.
All configurations are used for the analysis at zero density, while the
finite density analysis is performed every 10 trajectories.
Further details on the simulation parameters are given in
Ref.~\cite{BS05}.
The same data set is used to study the phase structure of two-flavor QCD
at finite density in Ref.~\cite{Ejiri:2007ga}. 

\begin{figure}[tb]
\begin{center}
\begin{tabular}{c}
\includegraphics[width=71mm]{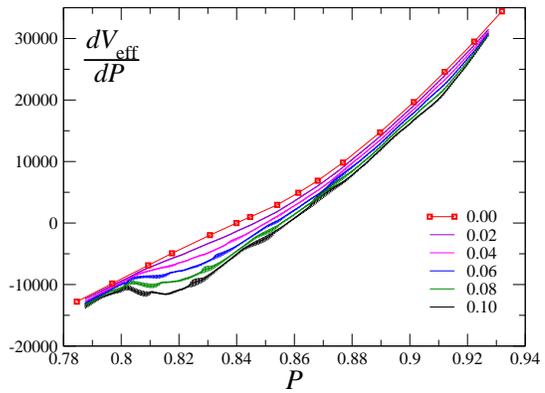}
\end{tabular}
\vspace{-3mm}
\caption{The 
slope of $V_{\rm eff} (P; \beta, h,0)$ 
normalized at $(\beta, h)=(3.65, 0)$ for $h=0.0$ -- $0.1$.
The squares are $dV_0/d P$.}
\label{fig:vslp}
\end{center}
\end{figure}

We first calculate the potential in two-flavor QCD at zero
density, $V_0(P; \beta)$, the first term in eq.~(\ref{eq:vefftrans}).
Because the finite temperature transition is crossover for two-flavor
QCD at a finite quark mass, the distribution function is
always Gaussian type.
We thus evaluate the curvature of 
$V_0$ using an identity for the Gaussian distribution,
$d^2 V_0/dP^2 = 6N_{\rm site}/\chi_P$,
where $\chi_P$ is the plaquette susceptibility, 
$\chi_P \equiv
6 N_{\rm site} \langle (P- \langle P \rangle )^2 \rangle$.
The slope of $V_0$ in the heavy quark limit can be also measured using
an equation derived from eqs.~(\ref{eq:vefftrans}) and (\ref{eq:lnr}). 
When one performs a simulation at $\beta_0$, the slope is zero at the
minimum of $V_0(P; \beta_0)$, and the minimum is realized 
at $P \approx \langle \hat{P} \rangle_{\beta_0}$. 
Hence, we obtain $dV_0/dP ( \langle \hat{P} \rangle_{\beta_0}, \beta) 
= - 6(\beta - \beta_0)N_{\rm site}$ \cite{ejiri09}.
The result of $d^2V_0/dP^2$ is plotted in the bottom panel of Fig.~\ref{fig:lnr}.
The circle symbols with dashed lines are calculated by $\chi_P$. 
The square symbols are computed by the numerical differential of 
$dV_0/dP$ obtained at the minimum of $V_0$. 
$dV_0/dP$ are the squares in Fig.~\ref{fig:vslp}.
These results obtained by two different methods are consistent.

\paragraph{Zero density.}
In the calculation of $\bar{R}(P; h,0)$, we use the delta
function approximated by
$\delta(x) \approx 1/(\Delta \sqrt{\pi}) \exp[-(x/\Delta)^2]$, 
where $\Delta=0.0025$ is adopted consulting the resolution and the statistical error. 
Because $\bar{R}(P; h,0)$ is independent of $\beta$, we mix all
data obtained at different $\beta$ as is done in
Ref.~\cite{Ejiri:2007ga}.
The results for $\ln \bar R(P;h,0)$ are shown by solid lines in
the top panel of Fig.~\ref{fig:lnr} for $h=0.01$ -- $0.07$. 
A rapid increase is observed around $P \sim 0.82$.
It is also important to note that the gradient becomes larger as $h$
increases.

The second derivative $d^2 \ln \bar R/dP^2$ is calculated by
fitting $\ln \bar R$ to a quadratic function of $P$ with a range of
$P \pm 0.015$ and repeating with various $P$.
The results are plotted in Fig.~\ref{fig:lnr} (bottom), 
where $d^2V_0/dP^2$ is also shown as the circles or the squares 
with dashed lines.
This figure shows that $d^2 (\ln \bar R)/dP^2$ becomes larger with
$h$, and the maximum around $P=0.81$ exceeds $d^2 V_0/dP^2$ for $h > 0.06$.
This indicates that the curvature of the effective potential,
$d^2 V_{\rm eff}/dP^2 =d^2 V_0 /dP^2 -d^2 (\ln \bar R)/dP^2$, vanishes
at $h\sim 0.06$ and a region of $P$ where the curvature is negative
appears for large $h$.
We estimated the critical value $h_c$ at which the minimum of
$d^2 V_{\rm eff}/dP^2$ vanishes and obtained $h_c=0.0614(69)$.

To see the appearance of the first order transition in a different way,
we plot $dV_{\rm eff}/dP$ at finite $h$ for $\beta^*=3.65$ in
Fig.~\ref{fig:vslp}.
The shape of the $dV_{\rm eff}/dP$ is independent of $\beta$ because 
$d^2 V_{\rm eff}/dP^2$ is $\beta$-independent.
$dV_{\rm eff}/dP$ is monotonically increasing when $h$ is small,
indicating that the transition is crossover.
However, the shape of $dV_{\rm eff}/dP$ turns into an S-shaped function
at $h \sim 0.06$, corresponding to the double-well potential.

We defined the parameter $h=2 \Nf \times (2\kh)^{N_t}$ for the
Wilson quark.
Then, the critical $\kappa_{hc}$ corresponding $h_c$ decreases as
$\kappa_{hc}=[h_c/(2\Nf)]^{1/N_t}/2$ with $\Nf$, and 
the truncation error from the higher order terms in $\kh$
becomes smaller as $\Nf$ increases.
The application range of the hopping parameter expansion was 
examined in quenched QCD simulations with $N_t=4$, by explicitly
measuring the size of the next-to-leading order (NLO) terms of the
expansion~\cite{whot12}.
They found that the NLO contribution becomes comparable to that in the
leading order at $\kh \sim 0.18$.
Hence, this method may be applicable up to around $\kh\sim 0.1$.
For instance, in the case of $\Nf=10$ with $N_t=4$, 
$\kappa_{hc}$ is 0.118.

\begin{figure}[tb]
\begin{center}
\begin{tabular}{c}
\includegraphics[width=35mm]{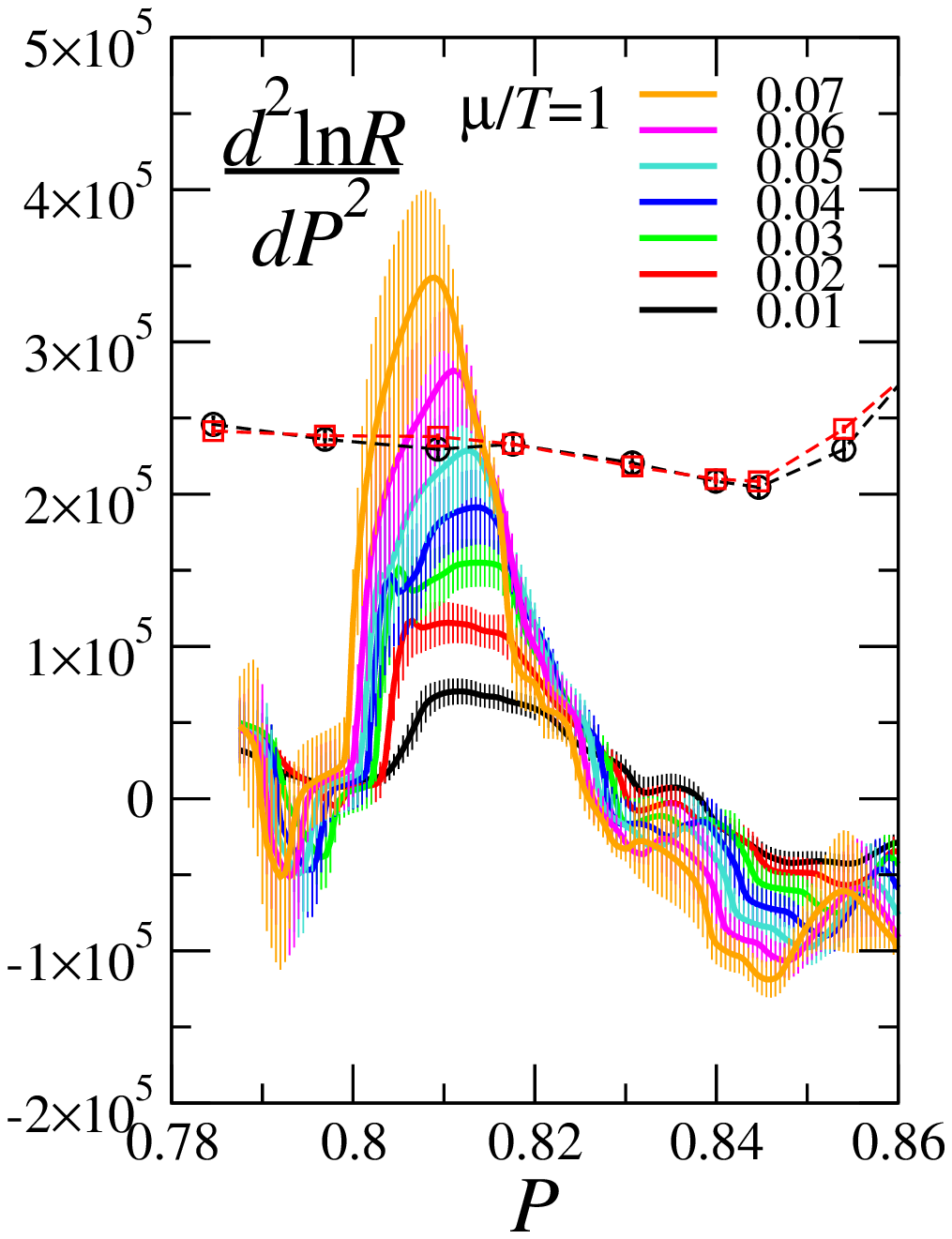}
\hspace{-2mm}
\includegraphics[width=50mm]{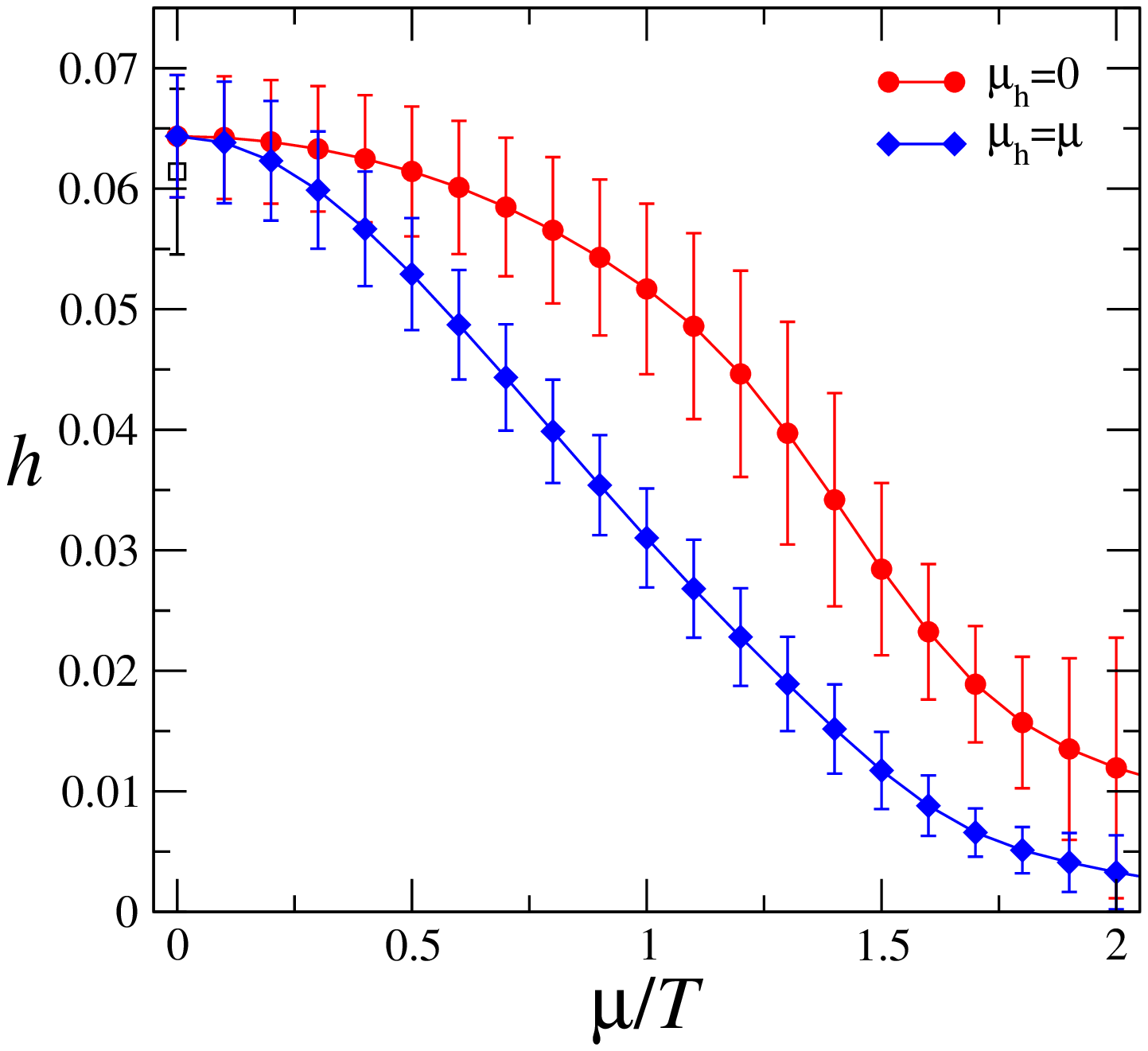}
\end{tabular}
\vspace{-3mm}
\caption{Left: The curvature of $\ln \bar{R} (P; h, \mu)$ 
as functions of the plaquette at $\mu /T=1.0$ and $\mu_{\rm h}=0$.
Right: The critical line in the $(h, \mu)$ plane for 
$\mu_{\rm h}=0$ (circles) and for $\mu_{\rm h}=\mu$ (diamonds). 
In the region above this line, the transition is of first order. 
The square at $\mu=0$ is computed using all configurations, 
and the others are measured with every 10.}
\label{fig:crtmu}
\end{center}
\end{figure}

\paragraph{Nonzero density.}
Finally, we turn on a chemical potential $\mu$ for two light quarks and 
$\mu_{\rm h}$ for $\Nf$ flavors, and 
discuss the $\mu$ dependence of the critical mass.
Because the strange chemical potential is small in 
the heavy-ion collisions, the $(2+1)$-flavor case with $\mu_{\rm h}=0$ 
corresponds to the experiments.
As discussed above, we can investigate the critical region easily for 
large $\Nf$.
$\bar R(P;h,\mu)$ is then given by  
$\langle (\det M(m_{\rm l},\mu)/ \det M(m_{\rm l},0))^2$ $\times$ 
$(\det M(m_{\rm h},\mu_{\rm h})/ \det M(\infty,0))^{\Nf} \rangle_{P: {\rm fixed}}$.
The quark determinant is computed using the Taylor expansion of
$\ln [\det M(m_{\rm l}, \mu)/ \det M(m_{\rm l}, 0)]$ in terms of 
$\mu/T$ up to $O[(\mu/T)^6]$ and the Gaussian
approximation is applied to avoid the sign problem as explained in
Ref.~\cite{Ejiri:2007ga}.
This approximation is valid for small $\mu$.
The truncation error has been estimated comparing 
the results up to $O(\mu^4)$ and $O(\mu^6)$ for $\mu/T \leq 2.5$ and 
is found to be small \cite{Ejiri:2007ga}.
The left panel of Fig.~\ref{fig:crtmu} shows the curvatures of $V_0$ and
$\ln \bar R(P;h,\mu)$ at $\mu/T=1$, $\mu_{\rm h}=0$. 
The maximum value of $d^2 \ln \bar R(P;h,\mu)/ dP^2$ is larger
than that at $\mu=0$.
This means the critical $h$ is smaller at finite $\mu$. 
Figure~\ref{fig:crtmu} (right) shows the critical value of $h$ 
as a function of $\mu$ for $\mu_{\rm h}=0$ (circles) and 
$\mu_{\rm h}=\mu$ (diamonds).
In the region above this line, the effective potential has the negative
curvature region, indicating the transition is of first order.
It is clear that the first order region becomes wider as $\mu$ increases.
If the same behavior is observed in $(2+1)$-flavor QCD, 
this gives the strong evidence for the existence of the critical point at
finite density in the real world.

Although this analysis is valid only for large $\Nf$, it gives a frame of
reference for the study of critical mass at finite $\mu$.
Notice that $\ln \bar{R}(P;h,\mu)$ is given by the sum of
$\ln \bar R(P;0,\mu)$ and $\ln \bar R(P;h,0)$ approximately and that
the behavior of $\ln \bar{R}(P;h,0)$ in Fig.~\ref{fig:lnr} is very similar 
to that of $\ln \bar{R}(P;0,\mu)$ in Figs.~5 and 7 in Ref.~\cite{Ejiri:2007ga}.
$\ln \bar R(P;0,\mu)$ is estimated from the quark number susceptibility
at small $\mu$ and $\ln \bar R(P;h,0)$ is obtained from the Polyakov
loop at small $\kh$.
Both the quark number susceptibility and the Polyakov loop rapidly
increase at the same value of $P$ near the transition point, which
enhances the curvature of $\ln \bar R$.
Therefore, the critical $h$ decreases with $\mu$ or 
equivalently the critical $\mu$ decreases with $h$.
The same argument is possible for $(2+1)$-flavor. 

\paragraph{\bf Conclusion and outlook.}
We studied the phase structure of (2+$\Nf$)-flavor QCD
to explore the realizability of the EW baryogenesis in technicolor
scenario and to understand properties of the finite density QCD.
Fixing the  mass of two light quarks,
we determined the critical mass of the other $\Nf$ quarks  
separating the first order and crossover regions.
The critical mass is found to become larger with $\Nf$.
Furthermore, the chemical potential dependence of the critical mass
is investigated for large $\Nf$, and the critical mass is found to
increase with $\mu$.

The next step for the estimation of the baryon number 
asymmetry in TC scenario is to quantify the strength of the first order
phase transition.
Another interesting application of our method is to study 
universal scaling behavior near the tricritical point. 
If the chiral phase transition in the two flavor massless limit is
of second order, the boundary of the first order
transition region $m_{\rm l}^c (m_{\rm h})$ is expected to behave as 
$m_{\rm l}^c \sim |m_h^{\rm tri.} -m_h|^{5/2}$
in the vicinity of the tricritical point, 
$(m_{\rm l}, m_{\rm h}, \mu)=(0, m_{\rm h}^{\rm tri.}, 0)$,
from the mean field analysis.
This power behavior is universal for any $\Nf$. 
The density dependence is important as well, which is expected to be
$m_{\rm l}^c \sim |\mu|^5$ \cite{ejirilat08}.
Starting from large $\Nf$, the systematic study of properties of 
QCD phase transition is possible.

\paragraph{Acknowledgments}
We would like to thank members of the WHOT-QCD Collaboration for discussions.
A part of this work was completed at the GGI workshop.
This work is in part supported by Grants-in-Aid of the Japanese Ministry
of Education, Culture, Sports, Science and Technology
(No.\
22740183, 
23540295 
)
and by the Grant-in-Aid for Scientific Research on Innovative Areas
(No.\
20105002, 
20105005, 
23105706  
).

\end{document}